# Simple non-invasive methods for obtaining the intensity and timing of arterial pulse waves

Ethan M. Rowland and Peter D. Weinberg

Department of Bioengineering, Imperial College London

## Abstract

Contraction of the left ventricle of the heart increases aortic root blood pressure ($P$), diameter ($D$) and blood velocity ($U$). When contraction diminishes, all three properties decrease. These perturbations propagate down the systemic arteries as the S wave and D wave, respectively. Peak carotid artery S-wave intensity is diminished and delayed in heart failure with reduced ejection fraction (HFrEF). A clinical trial demonstrated that these changes can be used to detect HFrEF with high sensitivity and specificity. Assessment of wave intensity and timing conventionally requires high-frequency, temporally and spatially coincident measurement of changes in $P$ and $U$ or $D$ and $U$ over the cardiac cycle. The practical difficulty of making such measurements accurately and noninvasively limits clinical utility. Here we test simpler methods by using numerical models of wave propagation and data from the clinical trial. We show that methods based on measuring only one of $P$, $D$ or $U$ can provide good surrogates for the full P-U and D-U methods. The best results were obtained when using measurement of $D$ to assess wave timing. That gave Receiver Operating Characteristics (ROCs) indistinguishable from those based on the full D-U method, with areas under the ROC of up to 0.905 when timing was anchored to the ECG rather than to other waves. Measuring vessel diameter over the cardiac cycle is technically simple and would be a cost-effective way of screening for HFrEF in primary care. Other metrics, similarly measured, might also allow screening for heart failure with preserved ejection fraction (HFpEF).

## Introduction

The left ventricle of the heart produces two forward-travelling pulse waves in the systemic arteries during each cardiac cycle [1]. The first, known as the S wave, is generated in early ventricular ejection. It is characterised by increases in vessel diameter, blood pressure and blood flow velocity. The second, known as the D wave, is generated by reduced ventricular contraction in late systole, and is characterised by decreases in the same three properties.

The waves may have diagnostic and prognostic value. For example, the S wave is reduced in intensity and delayed in compensated systolic heart failure [2, 3]. In a recent clinical trial [4], we showed that those changes are sufficiently sensitive and specific that they could be of use for detection of heart failure with reduced ejection fraction (HFrEF) in individual subjects. The timing of the wave was a better discriminator than its intensity.

Wave intensity, dI, can be determined by measuring changes over the cardiac cycle in pressure ($P$) or diameter ($D$) and velocity ($U$). The initial formulation of Parker and Jones [5] used $P$ and $U$:

$$dI = dPdU, \qquad (1)$$

where dP and dU are the change in pressure and velocity over a short interval. A later formulation by Feng and Khir [6] used $D$ and $U$:

$$_ndI = dDdU. \qquad (2)$$

We refer to these as the P-U method and D-U method, respectively. An equation in terms of vessel cross-sectional area ($A$) and volumetric flow rate ($Q$), analogous to equation (2), has also been used [7].

Ideally, the variables should be measured simultaneously, at the same arterial site, and with high temporal resolution. However, it is technically challenging to obtain accurate, non-invasive recordings that conform to these requirements. That limits the clinical utility of wave intensity analysis; simpler methods are needed.

Parker, Hughes and co-workers have developed a method that requires measurement only of $P$. Values are split into a reservoir pressure and an excess pressure using parameters obtained by assuming a mono-exponential decay during diastole [8,9]. Flow can then be calculated from the excess pressure [9] and that in turn allows calculation of wave intensity using the original P-U method [10]. We refer to this as the P-reservoir method.

Deriving flow velocity directly from excess pressure assumes an absence of reflected waves. In a trial of the method [11], calculated velocities correlated well with measured velocities in most subjects, but $R^2$ was <80% in 25% of them; the latter group had greater reflections. An additional issue is the need to anchor the excess pressure, and hence the velocity, since the range of values given by the calculations is free-floating. Previous work did this by assuming a maximum cross-sectionally averaged velocity of 1 m/s in the aorta [10], but that may be inappropriate for some patient groups; velocities are decreased in HFrEF [2], for example.

Our most accurate method for diagnosing HFrEF depends only on the timing of the S-wave and D-wave peaks in the carotid artery, relative to the ECG [4]. Wave intensities are needed to define the peaks but values do not have to absolute, nor are they required throughout the cardiac cycle or in other vessels. Since wave reflections are largely absent at these times in this vessel [2], we hypothesised that estimating intensity from $dP$, $dD$ or $dU$ alone, without an intermediate calculation of one of the other variables, would suffice. We term these methods P-only, D-only and U-only.

Here we describe a comparison of the P-reservoir and P-only methods with the conventional P-U method, and a comparison of the D-only and U-only methods with the conventional D-U method, using data from a previously published numerical model to avoid effects of measurement error and non-linear wall mechanics [12,13]. We also describe a comparison of the D-only and U-only methods with the conventional D-U method using data from our heart failure trial [4]. (Pressure was not measured in that trial.) In all cases, the focus was on the intensity and timing of the S-wave and D-wave peaks in the common carotid artery.

**Methods**

*1. Data*

Two previously published human carotid datasets were used. The first was generated from 1-D computational modelling using a range of cardiovascular properties derived from the healthy adult population to create 3,325 "virtual subjects" [12,13]. The model employed a time step, $dt$, of 1 ms and assumed perfectly elastic arterial walls. Single cycle $P$, $D$ and $U$ waveforms from the mid-point of the left carotid artery were selected. Since the model did not incorporate cardiac electrical activity, there was no ECG; the opening of the aortic valve was chosen as $t = 0$.

The second dataset was obtained from our clinical trial of HFrEF and control patients [4]. The HFrEF group had signs and symptoms of heart failure and a left ventricular ejection fraction (LVEF) < 40%, whilst the control group did not have signs and symptoms of heart failure and an LVEF $\geq$ 50%. Ensemble-averaged $D$ and $U$ waveforms were obtained by tracking blood and wall speckles at the mid-point of the carotid artery in a 6-second sequence of B-mode ultrasound images, with $dt$ = 1 or 1.3 ms. The peak of the R wave of a simultaneously acquired ECG was used as $t$ = 0. Sample sizes differed between wave metrics since an ECG could not be obtained in all subjects, and between left and right carotids since it was not possible to image both arteries in all subjects, but $n$ ranged from 32 to 47.

## 2. Calculation of wave intensity

The P-U method used

$$dI_{PU} = dPdU/(dt^2), \qquad (3)$$

a modification of the original formulation in equation (1) that removes the dependence of magnitudes on the time step [14].

For the same reason, the D-U method in equation (2) was replaced with:

$$dI_{DU} = dDdU/(dt^2). \qquad (4)$$

The methods of Parker, Hughes and co-workers [8-11] were used for the P-reservoir method. Briefly, an exponential decay was fitted to the pressure waveform during diastole using the form:

$$P_{res} - P_{zf} = (P_{res} - P_{zf})e^{-k_d t}, \qquad (5)$$

where $P_{res}$ is the reservoir pressure, $P_{zf}$ is the zero-flow pressure (i.e., the pressure at which outflow through the microcirculation ceases), and $k_d$ is the diastolic rate constant.

$P_{res}$ was calculated as:

$$P_{res} = e^{-(k_s+k_d)t} \int_0^t P(t')e^{(k_s+k_d)t'} dt' + \frac{k_d}{k_s+k_d} (1 - e^{(-k_s+k_d)t})P_{zf}, \qquad (6)$$

where $k_s$ is estimated by minimizing the squared error between $P$ and $P_{res}$ during diastole. Equations 5 and 6 were fitted to the full diastolic period, starting at maximum negative $dP/dt$. (Similar results were obtained when fitting only to the last two thirds of diastole).

Two scalings of the excess pressure, $P_{xs}$, were used to estimate $U$. Both rely on the fact that $U$ is known throughout the cardiac cycle in all virtual subjects in the numerical simulations. In the first, $P_{xs}$ was scaled separately for each subject:

$$U_{est} = P_{xs} (U_{max} - U_{min}) / U_{min}. \qquad (7)$$

$U_{min}$ is non-zero in the carotid. In the second, population average velocities were used:

$$U_{est} = P_{xs} (\overline{U_{max}} - \overline{U_{min}}) / \overline{U_{min}} \qquad (8)$$

where $\overline{U_{min}}$ and $\overline{U_{max}}$ are the dataset means. $U_{est}$ and $P_{xs}$ were used to calculate $dI_{Res}$ using equation (3).

The P-only analysis used:

$$dI_P = dP \tag{9}$$

and the D-only analysis used:

$$dI_D = dD. \tag{10}$$

### 3. Wave intensity analysis

Intensity waveforms were Savitzky-Golay filtered. The absolute intensity and timing of the S wave and D wave were determined at their peaks. The ECG-S interval was defined as the time between the ECG R-wave and arterial S-wave peaks, and the S-D interval was defined as the time between the arterial S-wave and D-wave peaks [refs: pre-clinical and clinical trials]. The S-wave shift (SWS) was defined as:

$$SWS = (60 \times ECG\text{-}S \text{ interval}) / (S\text{-}D \text{ interval} \times HR), \tag{11}$$

where HR is heart rate and the constant gives units of seconds [4].

For the numerical simulations, the ECG-S interval was replaced by the A-S interval, defined as the time between aortic valve opening and the S-wave peak.

### 4. Statistics

Spearman's rank correlation coefficient, $\rho_s$, was used to quantify correlations between data obtained using different wave intensity formulations. Bland-Altman plots were used to assess bias.

For the clinical dataset, Receiver Operator Characteristics (ROCs) were used to characterise the diagnostic potential of wave intensities and timings calculated by the D-only, U-only and D-U methods. The potential was quantified as the area under the ROC (AUROC). An AUROC of 0.5 indicates a purely random classifier and 1 indicates a perfect one; a value of 0.7 to 0.8 is considered acceptable, 0.8 to 0.9 is considered excellent, and more than 0.9 is considered outstanding [15].

## Results

*1. Wave intensities obtained with the P-reservoir method in virtual subjects*

S-wave peak intensity (SWI) and D-wave peak intensity (DWI) were determined by both the P-reservoir method and the "ground truth" P-U method in the left carotid artery of each virtual subject. The data from the two methods were then plotted against each other (Figure 1).

Linearity was poor, scatter was substantial and data lay far from the line of identity when scaling in each subject was by $\overline{U_{min}}$ and $\overline{U_{max}}$ – i.e., by velocities averaged across the population (Figure 1, top row).

Scaling by subject-specific $U_{min}$ and $U_{max}$ gave a more linear relation between the two methods with slope closer to 1 and much less scatter. Nevertheless, corresponding Bland-Altman plots (Figure 1, bottom row) showed significant deviations from a perfect agreement: confidence intervals for the mean difference were < 0 and there was an obvious trend in the data, indicating that P-reservoir data were consistently lower than the P-U data and that differences between the two methods were systematically related to the mean value obtained by both methods, indicating a proportional bias. (Correlation coefficients, $\rho_s$, were similar between the top row and centre row because they were calculated by Spearman's method, in which absolute values are replaced by their rank.)

Effects of individual variation in *dU* are included in the P-U method and in the P-reservoir method scaled by subject-specific $U_{min}$ and $U_{max}$, but not in the P-reservoir method scaled by population-average velocities $\overline{U_{min}}$ and $\overline{U_{max}}$. The fact that more nearly linear relations were obtained in the middle row of Figure 1 but than in the top row therefore indicates that there was covariance between the variables – that is, virtual subjects with larger values of *dP* also tended to have larger values of *dU.*

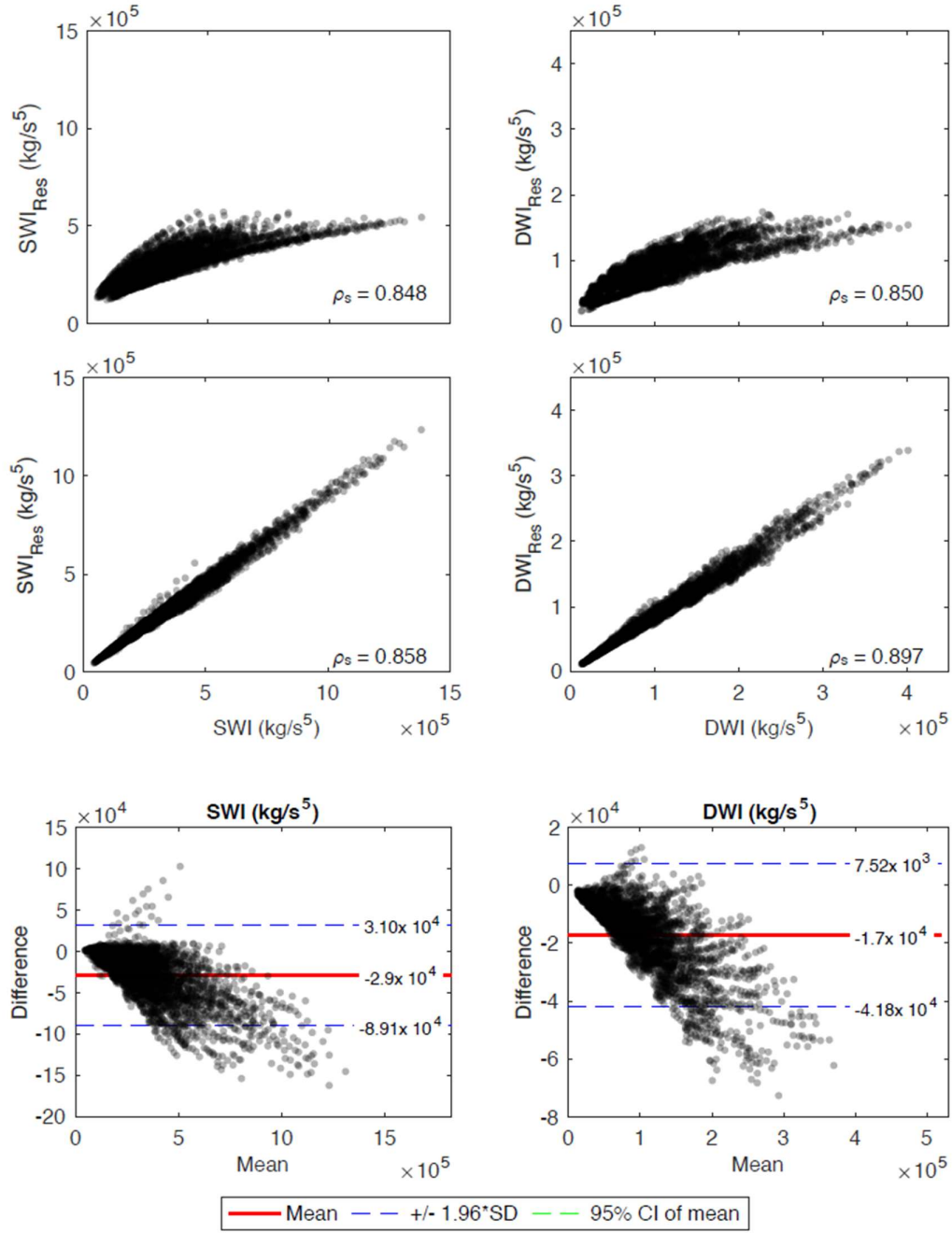

Fig 1. S-wave peak intensity (SWI; left column) and D-wave peak intensity (DWI; right column), obtained by applying the P-reservoir method (*y*-axis) or the P-U method (*x*-axis) to left carotid data from virtual subjects (*n* = 3,325). Scaling was by minimum and maximum velocities averaged across the whole population of virtual subjects (top row) or by subject-specific minimum and maximum velocities (middle row). Bland-Altman plots (bottom row) use the data from the top row. 95% confidence intervals are too close to the mean (red) line in the Bland Altman plots to be visible.

## 2. Wave intensities obtained with the P-only, D-only and U-only methods in virtual subjects

SWI and DWI were also determined in the carotid artery of each virtual subject by the P-only, D-only and U-only methods. Data obtained with the P-only method were plotted against "ground truth" data obtained with the P-U method, while data obtained with the D-only and U-only methods were plotted against "ground truth" data obtained with the D-U method (Figure 2).

Peak *dP* and peak *dD* were multiplied by the population average value of peak *dU* for the same wave, while peak *dU* was multiplied by the population average value of peak *dD* for the same wave. Note that such scaling does not influence the scatter, offset or linearity of the plots and would not be necessary in clinical applications of the single-variable methods. It was employed here only to make absolute values and units comparable between methods.

In all six comparisons, linearity was poor and data lay far from the line of identity. As with the top row of Figure 1, nonlinearity may have been caused by the use of population-average velocities for dP and dU, and population-average diameters for dU, since that does not take into account covariance of *dP* and *dU*, or *dD* and *dU*.

Non-linearity might be reduced by replacing *dD*, *dP* and *dU*, with $(dD)^2$, $(dP)^2$ and $(dU)^2$, respectively, rather than multiplying by population-average values. That would have the additional benefit of retaining the positive values of *dI* for the D wave seen with the two-variable methods. (D-wave intensities are negative when using *dP*, *dD* or *dU* alone).

Scatter was greater for the P-only method than for the D-only and U-only methods. That may have resulted from the use of a wide range in values of vessel stiffness in the simulations. *dU* and *dD* both reduce in stiffer vessels (the former due to reflections and the latter as a direct effect), whilst *dP* increases. Nevertheless, Spearman's rank correlation coefficients differed by <15% between the P-only method ($\rho \geq 0.863$) and the other two methods ($\rho \leq 0.990$).

The performance of the P-only method (Figure 2, top row) was visually similar to that of the P-reservoir method when scaling by population average velocities (Figure 1, top row).

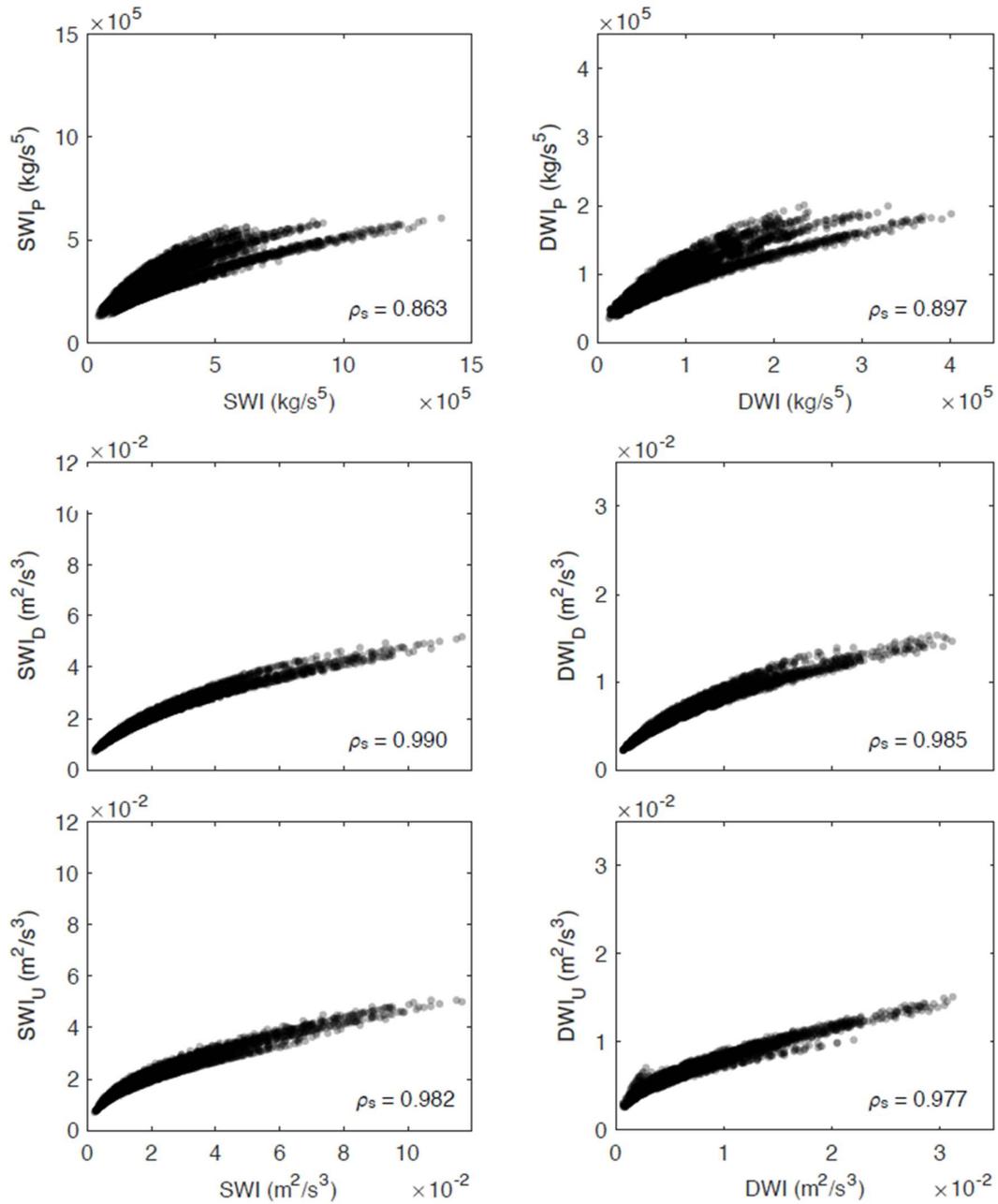

Fig 2. S-wave peak intensity (SWI; left column) and D-wave peak intensity (DWI; right column) in the left carotid artery of virtual subjects (*n* = 3325), obtained by applying the P-only method and the P-U method (top row), the D-only method and the D-U method (middle row) or the U-only method and the D-U method (bottom row); ; the first method of each pair is plotted on the *y*-axis and the second method on the *x*-axis. P-only and D-only results were multiplied by *dU* at the peak of the relevant wave and U-only results were multiplied by *dD* at the peak of the relevant wave, in all cases averaged across subjects, to obtain comparable values and consistent dimensions.

### 3. Wave timing obtained with the P-reservoir, P-only, D-only and U-only methods in virtual subjects

Intervals between aortic valve opening and peak S-wave intensity (the A-S interval) and between peak S-wave intensity and peak D-wave intensity (the S-D interval) were assessed in virtual subjects using P-reservoir, P-only, D-only and U-only methods to identify the wave peaks. Data obtained with the P-reservoir and P-only methods were compared with results from the P-U method, and data obtained with the D-only and U-only methods were compared with results from the D-U method (Figure 3). The quantisation of results arose because the range of A-S intervals and of S-D intervals was small compared to the time step used in the model.

In every case except one, scatter was small and linearity excellent. The exception was when using the U-only method to determine the S-D interval. In that case, there were two groups of points; both groups showed little variation in S-D interval estimated by the U-only method over a wide range of values obtained by the D-U method, and the nearly constant U-only value differed between the two groups. This behaviour is discussed below.

In all other cases, there was also a small (<10 ms) offset, discussed further in section (6) below.

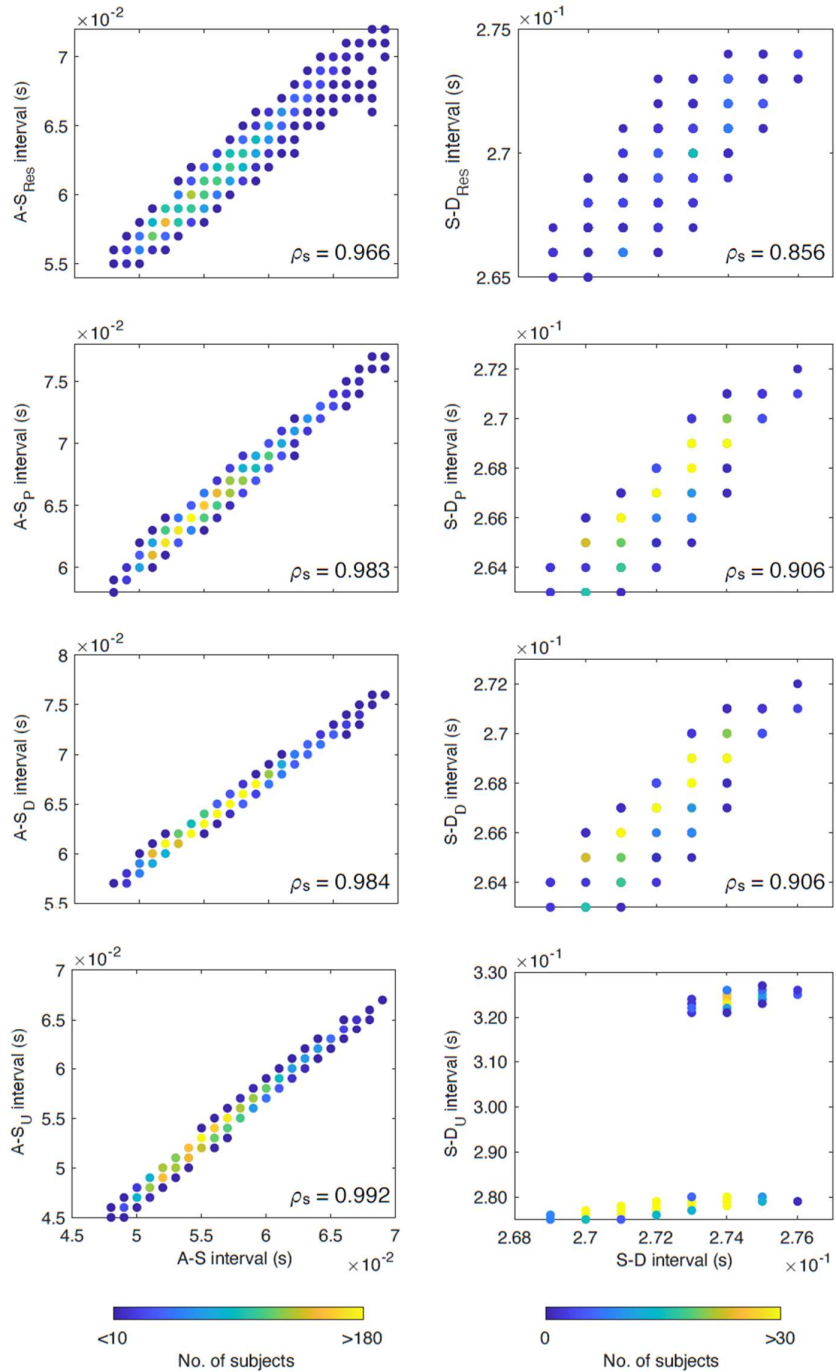

Figure 3. Time between aortic valve opening and the S-wave peak (A-S interval, left column) and between S-wave and D-wave peaks (S-D interval, right column) in the left carotid artery of virtual subjects ($n$ = 3325). Comparisons are between (from top to bottom) the P-reservoir method and the P-U method, the P-only method and the P-U method, the D-only method and the D-U method, and the U-only method and the D-U method; the first method of each pair is plotted on the y-axis and the second method on the x-axis. In each plot, many pairs of values occurred multiple times – the number of occurrences is indicated by colour.

*4. Wave intensities and timings obtained with the D-only method in patients*

Wave metrics were obtained in the left carotid artery of patients by the D-only method, and plotted against comparable data obtained by the D-U method. The D-only data were multiplied by the population average value of *dU* for the same metric to make the results directly comparable. ROCs and AUROCs were used to determine whether metrics determined by the D-only method was as good as metrics determined by the D-U method at discriminating between patients with and without HFrEF. Similar results were obtained for the right carotid artery (data not shown).

Plots comparing the D-only and D-U methods for both SWI and DWI showed a broadly linear trend with negligible offset (Figure 4 top row). Linearity was better than when using numerical data (compare Figure 4 top row with Figure 2 middle row), even though change in one variable was again plotted against the product of change in two variables. This indicates lower covariance of measured diameter and velocity in real subjects than in virtual ones and contradicts the idea that using $(dD)^2$ should preferred to *dD*. Covariance might have been lowered in vivo by homeostatic mechanisms that were not included in the numerical model, or by random errors in the measurement of *dU*.

Scatter for SWI and DWI was greater than for the equivalent numerical data. Again, this could reflect the lack of covariance – i.e., a given value of peak *dD* could be associated with a range of values of peak *dU*.

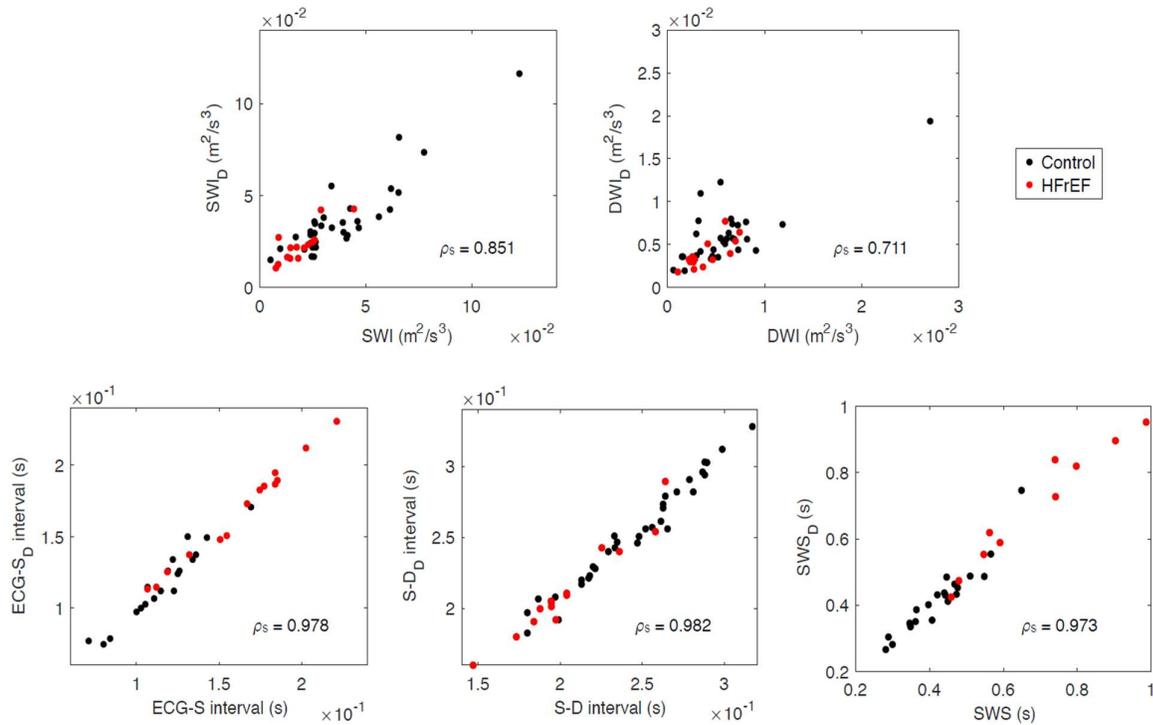

Figure 4. Top row. Peak S-wave and D-wave intensity (SWI, left, and DWI, right) in the left carotid artery of clinical trial patients, obtained using the D-only method (*y*-axis) or the D-U method (*x*-axis). D-only results were multiplied by *dU* at the peak of the relevant wave, averaged across subjects, to obtain comparable values and consistent dimensions. Bottom row. Equivalent plots for the interval between the R wave of the ECG and the S-wave peak (ECG-S interval, left), the interval between the S-wave and D-wave peaks (S-D interval, centre) and the S-wave shift (SWS, right). Red spots: HFrEF patients. Black spots: control patients. *n* = 47 except for ECG-S interval and SWS where *n* = 32.

Bland-Altman plots (Figure 5 top row) confirmed these trends: the 95% confidence intervals for the mean value included zero, indicating an absence of bias, and there was no obvious trend when examining the difference between methods as a function of the mean of both methods.

Equivalent comparisons of the D-only and D-U methods for the wave-timing metrics – the ECG-S interval, S-D interval and SWS – showed highly linear trends, close to the line of identity, in all three cases (Figure 4 bottom row). All three plots had negligible scatter: $\rho_s$ varied from 0.973 to 0.982. (Values of $\rho_s$ for SWI and DWI were lower, at 0.851 and 0.711, respectively.)

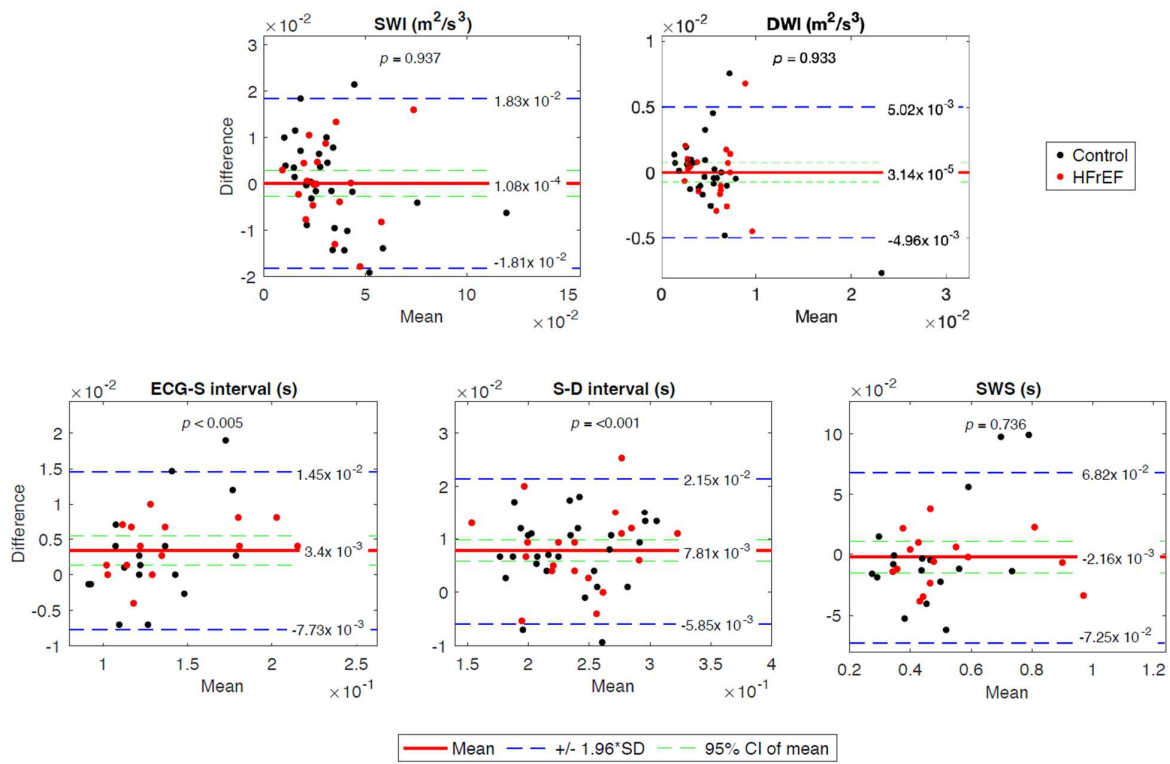

Figure 5. Bland-Altman plots (bottom row) use the data from Figure 4, arranged in the same order. *n* = 47 except for ECG-S interval and SWS where *n* = 32.

The corresponding Bland-Altman plots (Figure 5 bottom row) confirmed these trends: although the confidence intervals did not include zero for the ECG-S interval and S-D interval, and there is an indication of a proportional bias for the ECG-S interval and SWS, the differences are consistently much smaller than the means. That is not true of the equivalent plots for SWI and DWI.

The better performance of the D-only method when applied to timing metrics rather than intensity metrics shows that the peaks were correctly localised in time, even when the magnitudes of those peaks were not assessed with perfect accuracy.

The ROC and AUROC obtained for SWI when using the D-only method were only slightly inferior to those for the D-U method (0.773 vs 0.825) (Figure 6 top left). For DWI, the D-only method actually gave better discrimination (0.736 vs 0.639) (Figure 6 top right), perhaps because of the errors in measurements of *U* around the time of the D wave.

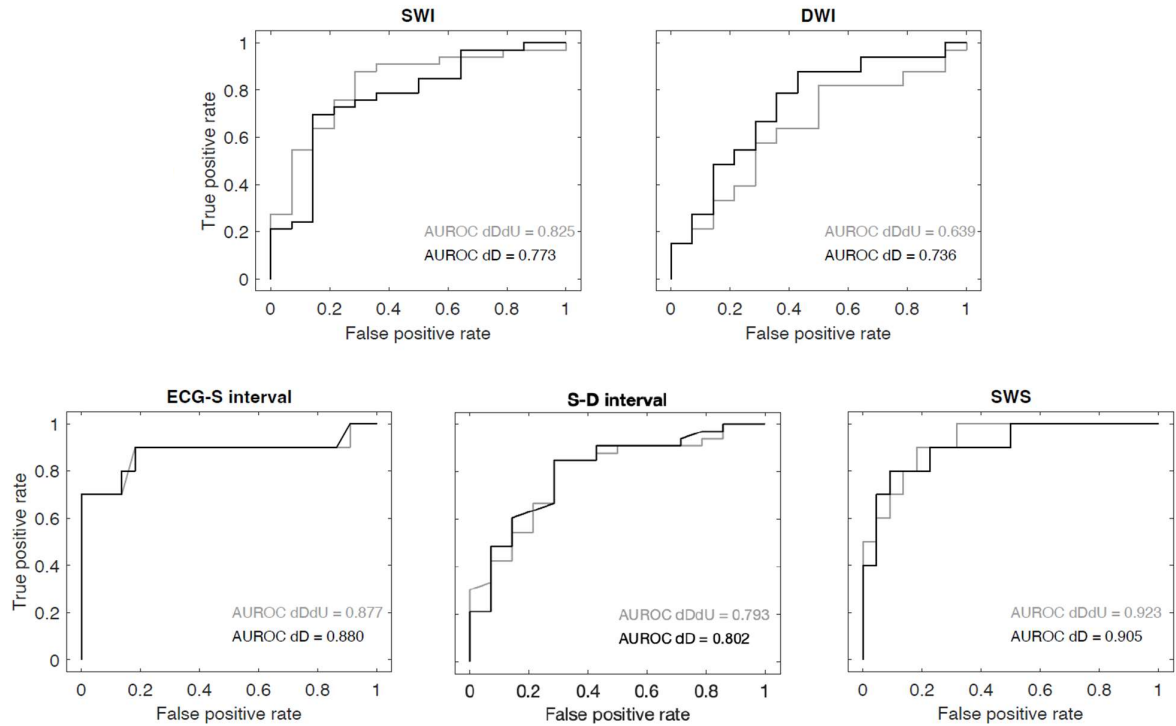

Figure 6. Top row. ROCs and AUROCs obtained for SWI and DWI, indicating the ability of these intensity metrics to distinguish between HFrEF and control patients when obtained by the D-only (black) or D-U (grey) methods. Bottom row. Equivalent ROCs and AUROCs obtained for the three timing metrics, the ECG-S interval, S-D interval and SWS. $n$ = 47 except for ECG-S interval and SWS where $n$ = 32.

For all three timing metrics, the ROCs and AUROCs obtained when using the D-only method were indistinguishable from those obtained when using the D-U method (Figure 6 bottom row).

## 5. Wave intensities and timings obtained with the U-only method in patients

SWI and DWI were also obtained in the left carotid artery of patients using the U-only method, and results were again plotted against comparable data obtained with the D-U method. The U-only data were multiplied by the population average value of *dD* for the same metric to make the results directly comparable. ROCs and AUROCs obtained with these reduced and "ground truth" methods were compared, as in the previous section (Figure 7). Similar results were obtained for the right carotid artery (data not shown).

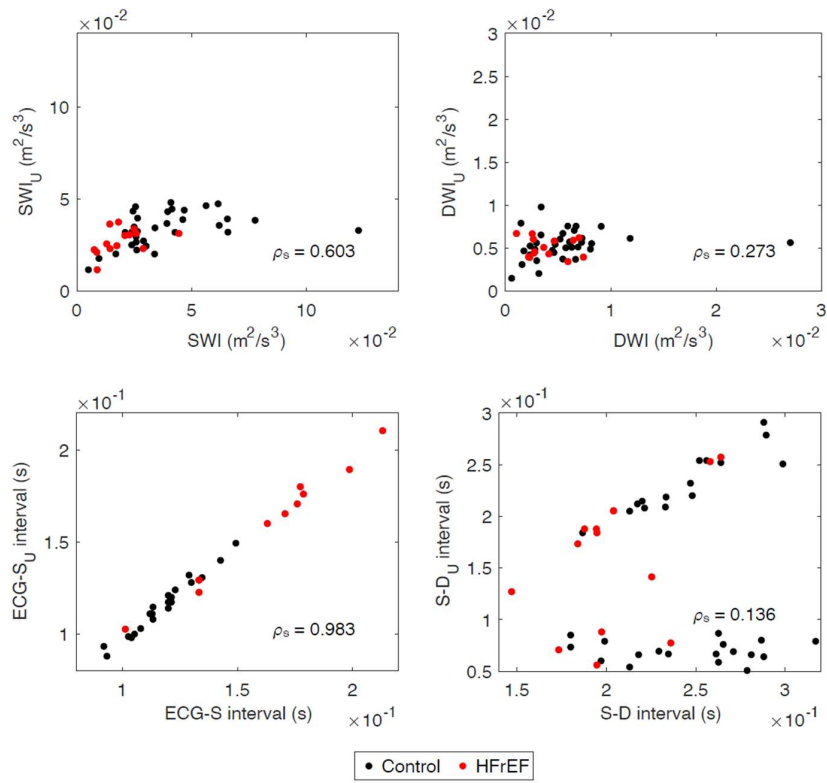

Figure 7. Top row. S-wave and D-wave intensity (SWI, left, and DWI, right, top row) in the left carotid artery of clinical trial patients, obtained using the U-only method (*y*-axis) or the D-U method (*x*-axis). U-only results were multiplied by *dD* at the peak of the relevant wave, averaged across subjects, to obtain comparable values and consistent dimensions. Bottom row. Equivalent plots for the interval between the R wave of the ECG and the S wave (ECG-S interval, left), the interval between the S wave and D wave (S-D interval, centre) and the S-wave shift (SWS, right). Red spots: HFrEF patients. Black spots: control patients. *n* = 47 except for ECG-S interval and SWS where *n* = 32.

Plots for both SWI and DWI again showed a broadly linear trend with negligible offset, but scatter was greater than for the comparison of D-only and D-U methods, especially for DWI, and there was a single extreme outlier for both metrics (Figure 8 top row): $\rho_s$ decreased from 0.851 to 0.603 for SWI, and from 0.711 to 0.273 for DWI.

The plot for the ECG-S interval again showed a linear trend with negligible offset (Figure 7, bottom left). $\rho_s$ was 0.983, which is essentially identical to the value obtained with the D-only method (0.978).

The plot for the S-D interval showed different behaviour (Figure 7 bottom right), similar but not identical to the aberration that occurred when the same method was used to obtain the same metric from the numerical simulation data. Approximately half the points showed a linear trend following the line of identity with similar scatter to the plot for the ECG-S interval, but the other half showed no discernible trend in the U-only data over the whole range of values obtained by the D-U only method. The phenomenon is discussed below. Because the U-only method gave obviously inconsistent results for the S-D interval, SWS – which incorporates the S-D interval – was not examined, and Bland-Altman plots were not employed.

ROCs and AUROCs obtained for SWI by the U-only method (Figure 8 top left) showed substantially lower discrimination than was obtained with the D-U method (0.716 vs 0.825). Again, this likely reflects errors when measuring $U$. The degradation was less for DWI (Figure 8 top right), but that may be because discrimination was poor even with the D-U method (U-only gave 0.569, while D-U gave 0.639).

For the ECG-S interval, the U-only method, like the D-only method, gave a ROC and AUROC that were indistinguishable from those obtained by the full D-U method (Figure 8 bottom left); the AUROC for the U-only method is thus comparable to that obtained by the D-only method (0.870 vs 0.880). The performance of the U-only method for the S-D interval (Figure 8 bottom right) was, however, much worse than that for the full D-U method (0.449 vs 0.793), as expected from the problems experienced with the former method when trying to estimate the S-D interval.

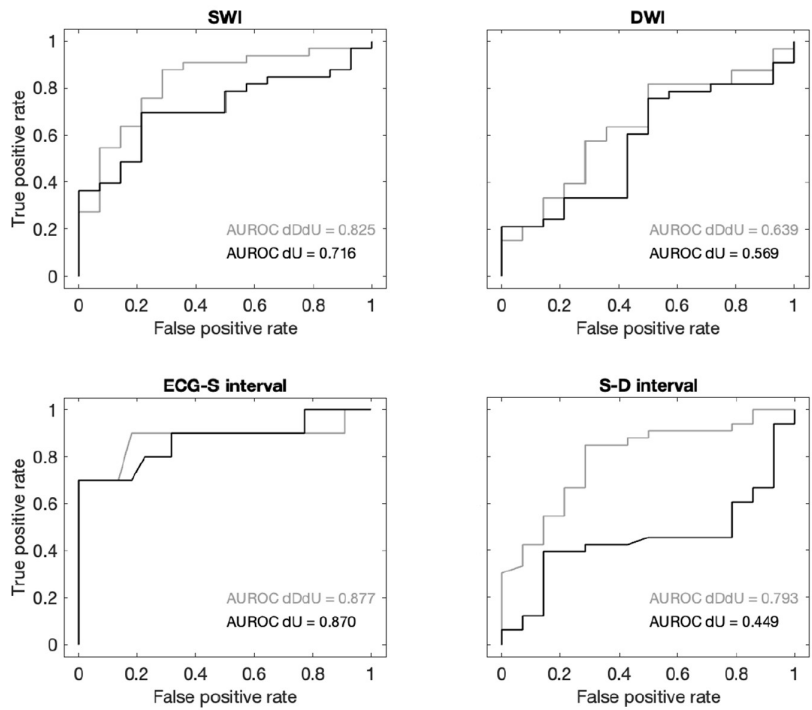

Figure 8. Top row. ROCs and AUROCs obtained by the U-only (black) or D-U (grey) methods for SWI and DWI, indicating the ability of these intensity metrics to distinguish between HFrEF and control patients when. Bottom row. Equivalent ROCs and AUROCs obtained for two timing metrics, the ECG-S interval and S-D interval. $n$ = 47 except for ECG-S interval where $n$ = 32.

## 6. Temporal coincidence of dD, dU and dDdU in virtual subjects and patients

*dD*, *dU* and *dDdU* in the carotid artery were plotted against time for a randomly chosen virtual subject (Figure 9 left) and real patient (Figure 9 right). Vertical lines indicate the occurrence of the S-wave and D-wave peak intensities, identified by the D-U method. *t* = 0 indicates aortic valve opening for the virtual subject and the peak of the ECG R wave for the patient.

*dP* was not included for the virtual patient – the data would be identical to the *dD* data since the model assumed purely elastic vessel walls – and was not included for the patient because *P* was not measured in the clinical trial. So that all three traces could be included on the same graph, values of *dD*, *dU* and *dDdU* were normalised to have a minimum of 0 and a maximum of 1. Without normalisation, peak D-wave values were always negative for *dD* and *dU*, and therefore positive for their product, *dDdU*; that inversion of the *dDdU* peak relative to the other two is visible on the graphs

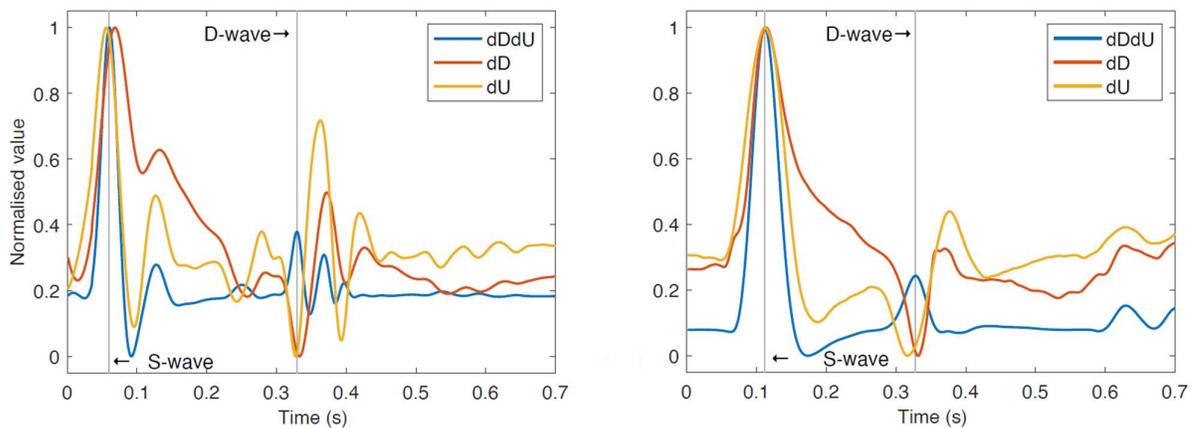

Figure 9. Plots of *dD*, *dU* and *dDdU* against time over a single cardiac cycle for a virtual subject (left) and a patient (right). All curves are normalised so that their highest value (S-wave peak) has a value of 1 and their lowest point (D-wave peak) has a value of zero. *t* = 0 indicates aortic valve opening for the virtual subject and the peak of the ECG R wave for the patient.

The curves for the virtual subject are in reasonable agreement with the curves for the patient, but more reflections are visible for the virtual patient. For example, the virtual patient has a visible notch in the dicrotic limb of the *dDdU* curve and a second *dDdU* peak after the D-wave.

In both graphs, S-wave and D-wave peaks in *dD* and *dU* occurred at the same time to a good approximation, but there were discrepancies. In the virtual subject, the S-wave peak in *dU* occurred visibly earlier than the corresponding peak in *dD*, a discrepancy that was not seen in the real patient. That is a further indication of exaggerated wave reflections in the former; they appear to occur early in the cardiac cycle and retard the flow of blood. An earlier peak in *dD* than in *dU* also occurred at the time of the D-wave, but in this case was greater in the real patient. Such discrepances likely account for the offsets seen when comparing D-only and D-U estimates of the A-S interval in virtual subjects (Figure 3 third row) and when making similar comparisons for the S-D interval in patients (Figure 5 bottom row). The offsets are acceptably small compared to the mean difference in ECG-S interval or S-D interval between HFrEF and control patients, which are both ≈50 ms.

## Discussion

The present study investigated the accuracy of estimating pulse wave intensity and timing by four reduced methods. They were the P-reservoir method of Parker, Hughes and co-workers [8-11], and the P-only, D-only and U-only methods proposed here. (There does not appear to be a theoretical justification for a D-reservoir method [R.M. Reavette, unpublished].) The full P-U method of Parker and Jones [5] and the full D-U method of Feng and Khir [6] were used as ground truth. All methods were tested on the dataset obtained by numerical modelling [12,13]. The D-only and U-only methods were additionally tested on data from our recent clinical trial [4], where $D$ and $U$ but not $P$ were measured, using the D-U method as ground truth.

Our motivation was the development of simple methods for diagnosing HFrEF. Our clinical trial [4] showed that wave metrics provide sufficient sensitivity and specificity for this purpose, but the need to measure arterial blood velocities with high accuracy and temporal resolution adds complexity: time-consuming, data-intensive spatiotemporal filtering is needed in order to separate the weak blood signal from the strong background clutter; tracking speckles in blood using autocorrelation methods on successive frames is computationally demanding; and an ultrafast scanner with plane wave transmission is required to capture the high velocities occurring in large arteries. There are likely to be other clinical and non-clinical applications of simpler methods, should they provide sufficient accuracy.

We first consider S-wave and D-wave intensities obtained using the P-reservoir method in conjunction with the dataset derived from numerical modelling. They were in excellent agreement with intensities obtained by the full P-U method so long as the results were scaled by velocities occurring at the same site in the same virtual subject. However, if individual velocity information were available in real pre-clinical or clinical datasets, there would be no need for a reduced method anyway. When population average velocity values, $\overline{U_{min}}$ and $\overline{U_{max}}$ were used, the P-reservoir results were monotonically but not linearly related to intensities obtained by the P-U method, and scatter was increased. Choosing other constant values of $U_{min}$ and $U_{max}$ would simply result in a linear scaling of the results for all subjects and would not affect the conclusion; scatter arises because a single assumed velocity may be more or less accurate for different people. Errors could theoretically also have resulted from choosing a suboptimal region of the diastolic decay for exponential fitting when partitioning measured pressure into its reservoir and excess components, given the assumption that it should be a period without wave reflections. We found the same trends when using two different regions, but other fitting strategies could be attempted.

The P-only and D-only methods avoid the need to scale the data and the need to fit a mono-exponential decay to diastolic pressure. Nevertheless, a very similar result was obtained when applying the P-only method to data from the numerical model, suggesting that it is also not a reliable way of obtaining precise and accurate wave intensities. The results from the D-only analysis were, again, monotonically but not linearly related to data from the full method, but scatter was much reduced. Hence the D-only method might be suitable for estimating S-wave and D-wave intensities, provided that only the order of intensities rather than their absolute value is required; it would be possible to say that one value was higher or lower than another, but not by what fraction.

The situation was different when assessing the ability of the P-based and D-based methods to determine wave timing. All three methods gave a straight-line relation with the "gold standard" P-U or D-U methods, although there were small offsets. Furthermore, the scatter was low. The offsets were approximately 10 ms, and the scatter was similar in magnitude. Hence the results support the use of all the reduced methods to estimate the timing of the S-wave and D-wave peaks. That is because finding the peak of each wave depends only on having a monotonic relation between estimated and true wave intensities.

The U-only method gave mixed results when applied to the model data. Accuracy was high when estimating S-wave intensity and timing, but not when characterising equivalent properties of the D wave. Wave intensity plots obtained by using the *dU* surrogate had a single, unambiguous early maximum corresponding the S wave, but multiple negative peaks, corresponding to the D wave and to subsequent decelerations caused by wave reflections. The negative peaks caused by reflections were larger in many cases and hence were erroneously characterised as the D wave by our algorithm. (That problem did not affect *dI* obtained by the full *dDdU* or *dPdU* methods because *dD* and *dP* were generally close to zero at the time wave reflections occur, but not when the D wave occurred.) The problem could probably be reduced but not eliminated by discounting negative peaks occurring outside of a time window defined relative to the S-wave peak or the R wave of the ECG, within which the D wave is expected to occur.

Finally, we consider the findings obtained when using data from the clinical trial. When the D-only method was used to estimate wave intensities, the relation with data obtained using the D-U model was linear and offsets were negligible. However, there was some scatter, more so for the D wave than the S wave. The ROCs obtained with the D-only method were similar to those from the D-U method; AUROCs were 6% lower and 2% greater for the SWI in left and right carotids, respectively, and 16% greater and identical for the DWI.

The agreement was even better when using the D-only method to assess wave timing: the trend was linear, there were no offsets, and scatter was low. The ROCs and AUROCs were essentially identical to those obtained with the D-U method. The U-only method gave mixed results similar to those obtained with the data derived from numerical modelling.

*Limitations and Assumptions*

We have used wave intensities obtained with the full D-U and P-U methods as ground truth. However, these methods are themselves already simplifications. For example, they assume constant pressure and velocity over the cross section and a constant wave speed throughout the cardiac cycle at each location. The numerical implementation assumed a purely elastic tube law.

The P-reservoir method assumes that the timescale for wave travel is much smaller than the timescale for changes in the $P_{res}$ waveform, that there are no reflections in early systole which would affect flow into the aortic root, and that there is a mono-exponential decline of pressure in diastole. Furthermore, $P_{xs}$ is related to volumetric flow rather than velocity, as used here, and a scaling factor is required in order to relate the two in absolute terms. Finally, although the method has previously been used in the carotid [8], the original derivation and supporting data concern the proximal aorta.

The P-only and D-only methods assume that the location of the S-wave and D-wave peaks can be determined using a period of the waves that is short compared to the period over which $P_{res}$ changes substantially, an assumption that is supported by previously published data (e.g. [10]). That obviates the need to separate reservoir and excess values because changes in reservoir pressure will not affect the timing or magnitude of the two peaks, although it will likely affect their relative magnitudes.

Furthermore, the fact that we are interested specifically in the peaks means that the analysis is sensitive to reflected waves only at the time of these peaks in the carotid. Again, previous data shows this assumption to be valid, both in normal subjects and HFrEF patients [2].

Finally, the D-only method assumes the vessel wall obeys an elastic tube law, since viscous effects would alter the relative timings of *dD* and *dU*. Our data show that to be an excellent approximation. The method does not require peak *D* and peak *U* to coincide; that is clearly not the case in the human carotid [16].

**Conclusion**

Clinical trial data show that the D-only method identifies the interval between the R wave of the ECG and the carotid S wave, and the interval between the carotid S wave and D wave, with high accuracy. ROCs show that timing metrics based only on diameter measurements have excellent potential for diagnosing HFrEF. The SWS metric gave the highest AUROC, whilst the S-D interval would permit HF screening without collecting an ECG. Because the reduced methods are only a surrogate for corresponding methods of full wave intensity analysis – even the dimensions they produce are different – and because intensities are primarily used to identify wave timing, we propose they are collectively described by the term Pseudo Wave Timing Analysis (PWTA).

A reanalysis of data from our clinical trial shows that HFrEF was characterised by a flattening of the D wave, parameterised as the ratio of its peak intensity (DWI) to the area underneath it (the D-wave energy, DWE) (Figure 10), as well as by a delayed S wave; our preliminary data show that patients having heart failure with preserved ejection fraction (HFpEF) show the former but not the latter (Figure 10). If confirmed by larger studies, that combination would permit identification of both HFrEF and HFpEF, and discrimination between them. We speculate that there is a sufficiently long reflection-free period around time of peak D-wave intensity that the shape of the D wave can also be characterised by the D-only method, and by other reduced methods presented here.

The carotid artery was chosen because of its proximity to the heart (which reduces the importance of variation in vasoconstriction and hence wave speed) and because there are no significant reflections at the time of the S wave and D wave, but other vessels may also be suitable. Similarly, features of the ECG other than the R-wave peak could be used as a timing datum. Measurements of *D* were made by ultrasound, but a range of techniques could be used including MRI for *D*, *U*, *A* or *Q*, ultrasound for *U*, and oscillometric cuffs, tonometers, strain gauges and photoplethysmography for *D*.

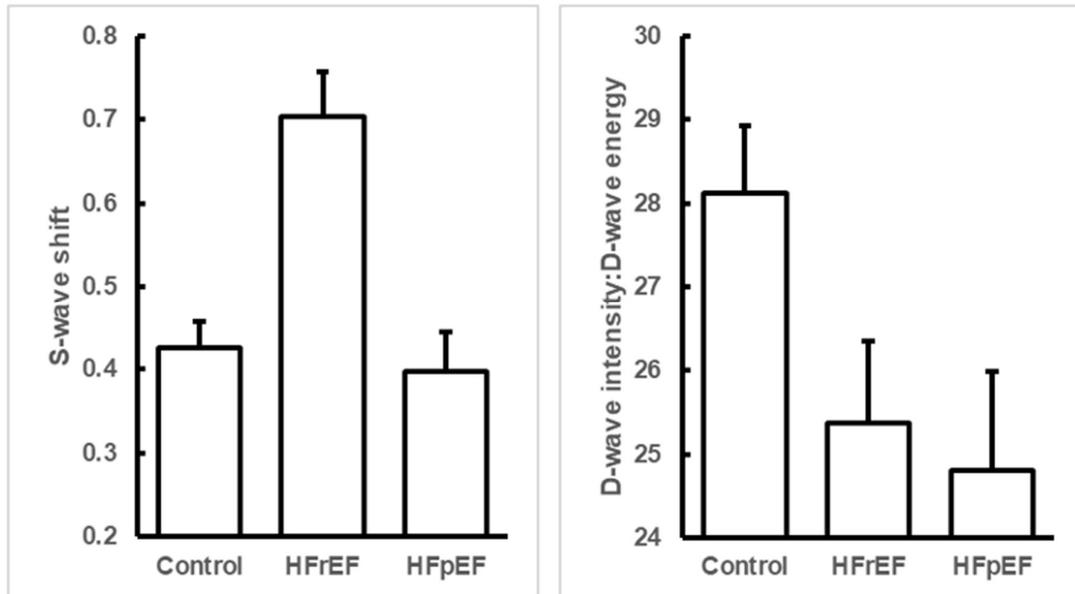

Figure 10. Left: S-wave shift can distinguish HFrEF ($n$ = 14) from both control ($n$ = 19; $p$ = 0.0002) and HFpEF ($n$ = 4, $p$ = 0.0014) patients. Right: HFrEF ($n$ = 14, $p$ = 0.004) and HFrEF ($n$ = 4, $p$ = 0.057) can be distinguished from control ($n$ = 33) patients by flattening of the D-wave, parameterised as the ratio of D-wave intensity to D-wave energy. Using both metrics together thus allows separation of HFrEF, HFpEF and control. Means±SEM

## Acknowledgements and Conflicts

Funded in part by a UKRI Impact Acceleration Award. EMR and PDW are inventors on a UK patent filing related to the data presented above.